\documentclass[12pt]{article}
\usepackage{graphicx}
\title{\bf X-raying the super star clusters\\
 in the Galactic center}
\author{L.~M.~Oskinova \\
\vspace{1cm}\\
\normalsize Astrophysik, Universit{\"a}t Potsdam, Germany}
\date{\mbox{}}
\begin{document}
\maketitle
\pagestyle{empty}
%
%
\def\bull{\vrule height .9ex width .8ex depth -.1ex}
\makeatletter
\def\ps@plain{\let\@mkboth\gobbletwo
\def\@oddhead{}\def\@oddfoot{\hfil\tiny\bull\quad
``Massive Stars and High-Energy Emission in OB Associations'';
JENAM 2005 'Distant Worlds' \quad\bull}%
\def\@evenhead{}\let\@evenfoot\@oddfoot}
\makeatother
%
%
\def\beginrefer{\section*{References}%
\begin{quotation}\mbox{}\par}
\def\refer#1\par{{\setlength{\parindent}{-\leftmargin}\indent#1\par}}
\def\endrefer{\end{quotation}}
%
%
{\noindent\small{\bf Abstract:} 

The Galactic center harbors some of the most massive star clusters known
in  the Galaxy: the Arches and the Quintuplet. Based on the Chandra 
observations of these clusters (PI: Wang) which recently became public,  
I discuss the X-ray emission from the massive stars in these clusters.
Confirming  the general trend for  Wolf-Rayet (WR) stars being X-ray dim, 
none of them is detected in the Quintuplet cluster. The most massive 
star known in the Galaxy, the Pistol star, is also  not  detected, invoking 
questions regarding the proposed binary nature of this  object. X-ray emission 
in the Arches cluster is dominated by three stellar point sources. All three 
sources as well as the cluster's diffuse radiation show strong  
emission at 6.4--6.7\,keV,  indicating the presence of fluorescenting cool 
material. The Arches point sources may be identified as colliding wind binaries, 
albeit other possibilities cannot be  ruled out. } 


\vspace{2cm}  

The X-ray emission from young stellar clusters, such as the Arches and the 
Quintuplet,  is tightly coupled with the massive stars evolution. 
When a massive star evolves from O type to WN type, and finally
to WC type, it keeps  its  bolometric luminosity nearly constant
while its X-ray  luminosity is likely to decline
(Oskinova\,2005). This trend is confirmed by  the study of WR stars in
the Galaxy and the LMC (Guerrero et al.\,in prep.,  also these
proceedings). In the 
1--3 Myr old Arches cluster (Figer et al.\,2002, F02), the majority of massive 
stars are of O type. The most massive stars are the most X-ray luminous ones 
due to the $\log L_{\rm X}=\log L_{\rm bol}-7$ correlation. The stellar wind 
input into the intracluster medium is expected to result in diffuse 
cluster wind emission. The level of cluster wind emission scales with stellar 
mass-loss rates and wind kinetic energy (Chevalier \& Clegg\,1985). Both 
quantities rise sharply when stars evolve to the WR phase, characterized by 
conspicuously powerful stellar winds. Therefore, in the 3--5 Myr old 
Quintuplet cluster (Figer et al.\,1999, FMM99) where a significant fraction 
of massive stars has evolved to the WR stage the level of cluster wind 
emission is expected to rise. The same time the stellar X-ray emission 
becomes much less prominent with most massive stars evolving fast to the
WC stage and becoming X-ray dim. 

%
%
  \begin{figure}
  \begin{minipage}{8cm}
  \centering
  \includegraphics[width=7cm]{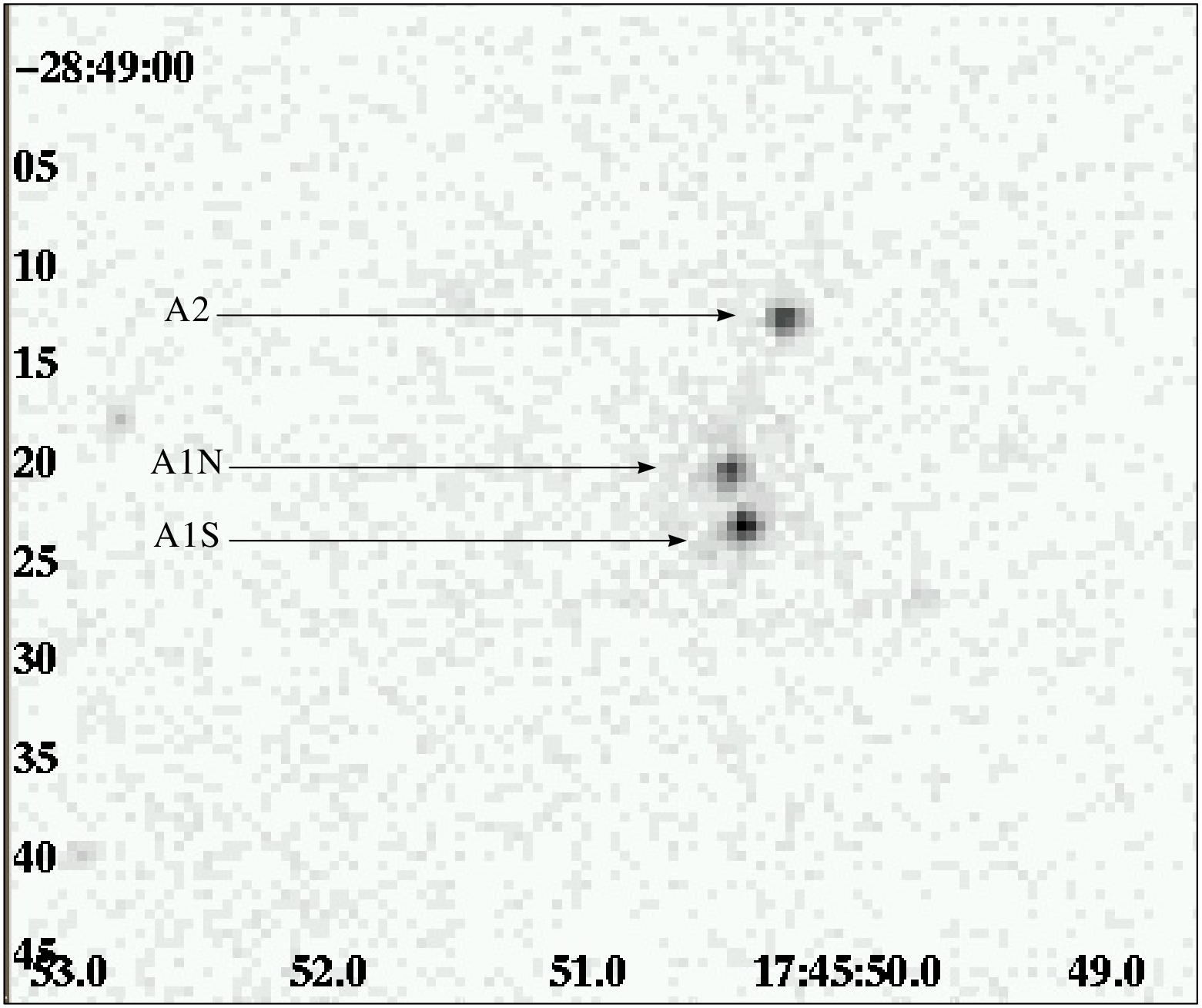} 
  \caption{ACIS-I image of the Arches. 
The emission is dominated by three point sources, as indicated by arrows. Names of 
the sources are the same as in Law \& Yusef-Zadeh\,(2004).} 
  \end{minipage}
  \hfill
  \begin{minipage}{8cm}
  \centering
  \includegraphics[width=7cm]{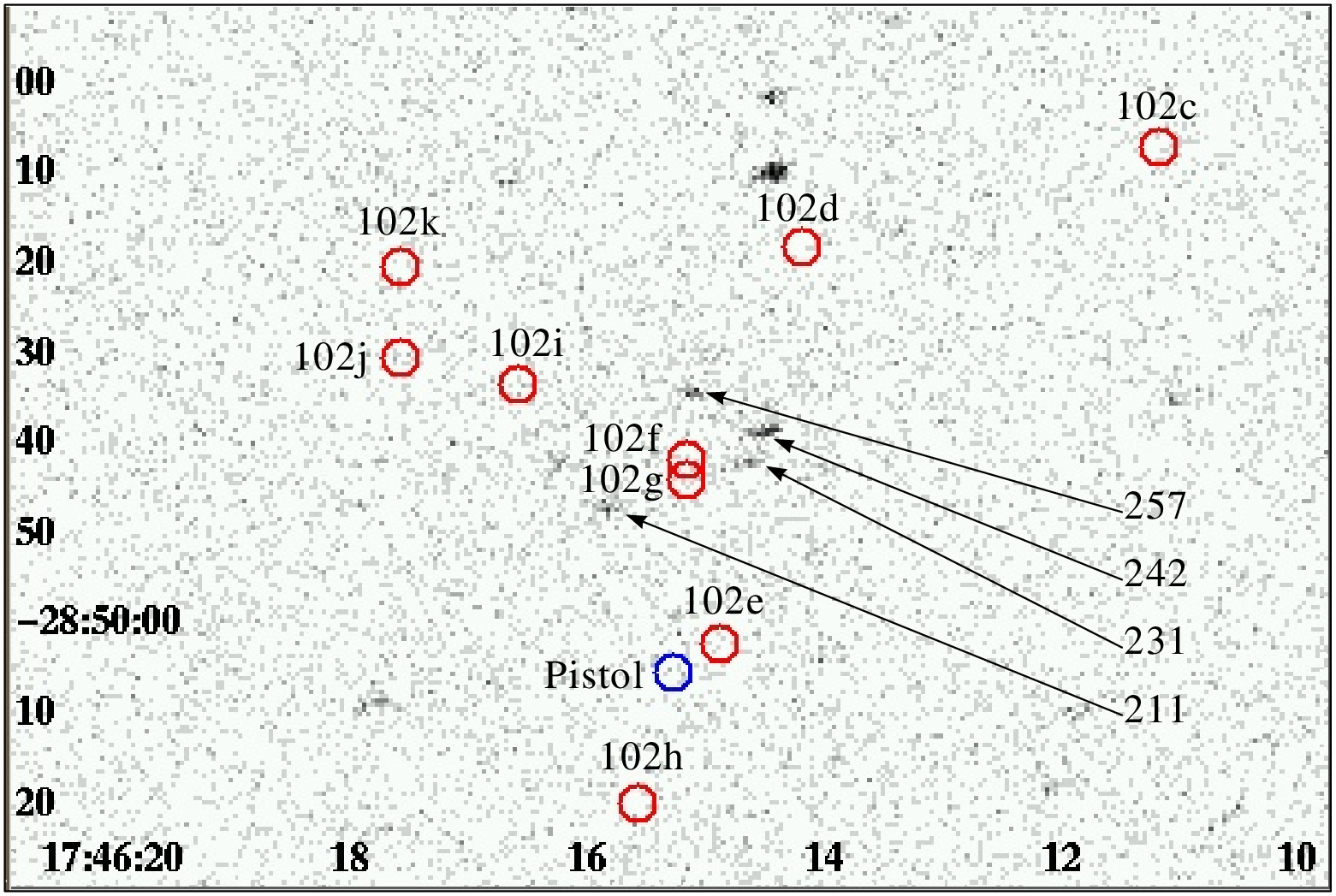} 
\caption{ACIS-I image of the Quintuplet. Positions of known WR stars 
(WR102c ... WR102k) and of the Pistol star are encircled. None of these
is detected. Detected sources with known IR counterparts are indicated by  
arrows (object numbers  are from  FMM\,99).}
  \end{minipage}
  \end{figure}
\vfill
The Arches and the Quintuplet were recently observed with Chandra in a 
nearly 100\,ksec exposure. Figure\,1 shows the ACIS-I image of the Arches 
cluster. Three bright prominent point sources (see Table~1) coincide with 
the stars F02-6~(A1S), F02-7~(A1N) and F02-9~(A2) (numbers from 
Figer et al.\,2002). F02-9 is an emission line star. The luminosities listed 
in the Table~1 are not corrected for absorption (the corrected $L_{\rm X}$ is 
$\approx 4$ times higher).  All three 
sources are suggested to be colliding wind binaries (Law \& Yusef-Zadeh\,2004). 
Their X-ray luminosity, however high, is not unique among Galactic binaries 
(e.g. HD\,150136 has $L_{\rm X}\approx 3\times 10^{33}$\,erg/s, 
Skinner et al.\,2005). It is their X-ray spectra that are rather unusual 
(Figs. 3,\,4).  
%
%
  \begin{figure}
  \begin{minipage}{8cm}
  \centering
  \includegraphics[width=7cm]{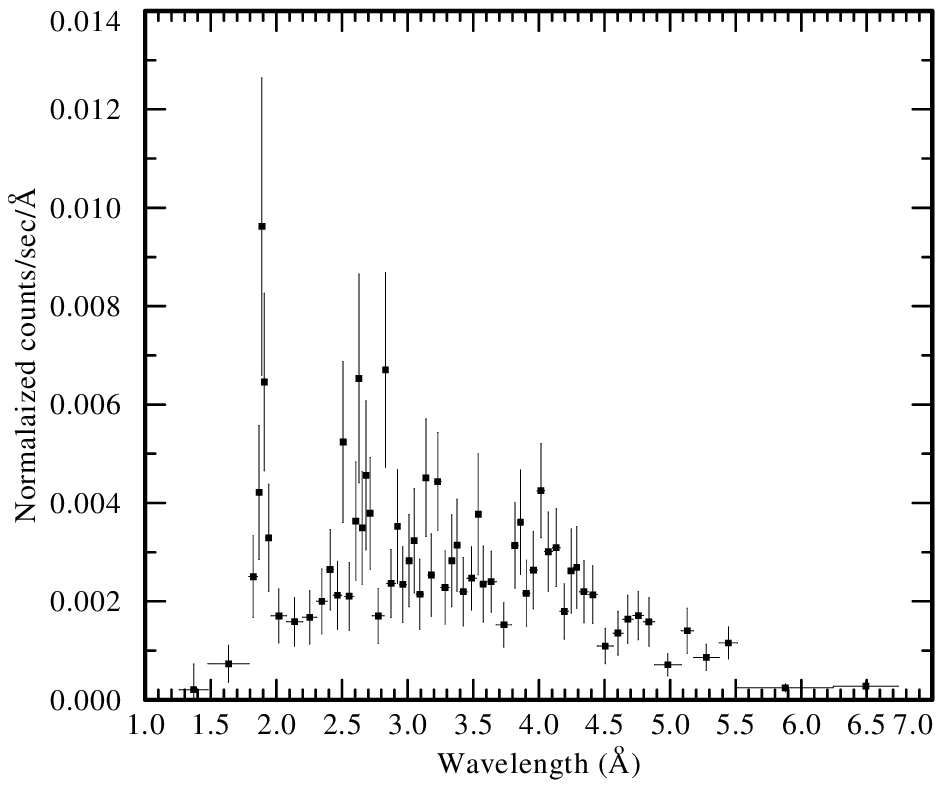} 
  \caption{Observed background-subtracted spectrum of the brightest X-ray 
point source A1S (FMM\,6) in the Arches cluster.} 
  \end{minipage}
  \hfill
  \begin{minipage}{8cm}
  \centering
  \includegraphics[width=7cm]{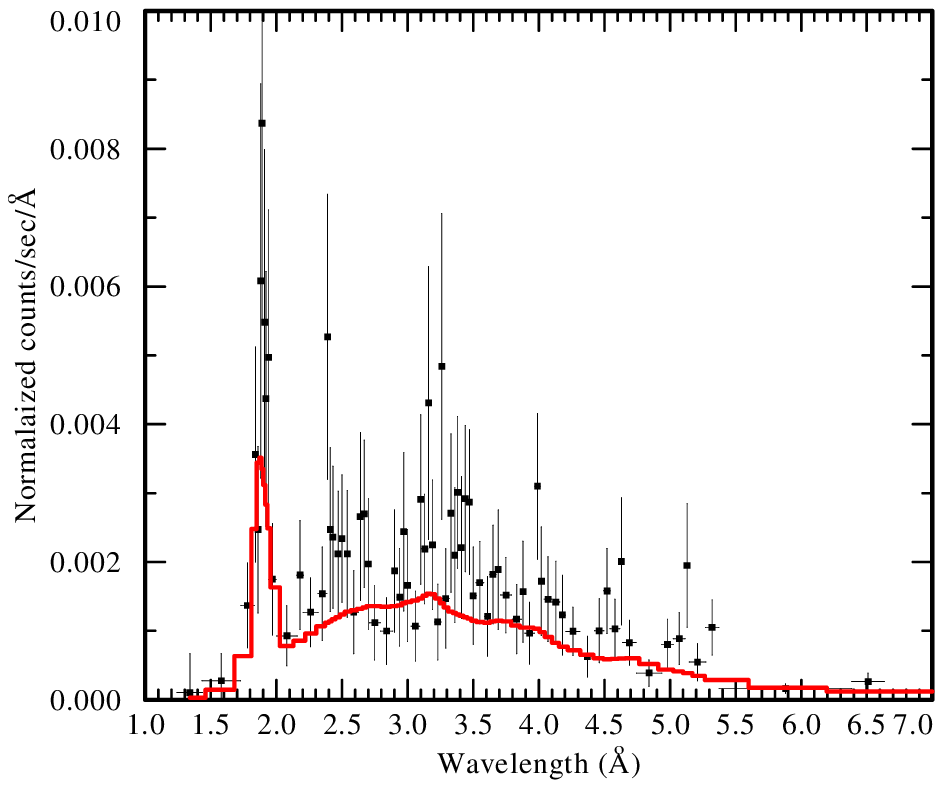}
  \caption{Observed background-subtracted spectrum of emission line
star A2 (FMM\,9). The solid line is a fitted  thermal plasma model with 
$kT=3.5\pm 1.2$\,keV with additional gaussian component at 1.7-2.1\AA.}
  \end{minipage}
  \end{figure}

Spectra of all three stars are remarkably similar in appearance, 
strongly  resembling the spectrum of $\gamma^2$~Vel in the low state
(Schild et al.\,2004). The highly absorbed spectra  of A1S, A1N and A2
can be fitted by a variety  of models, all of them needing an
additional component describing the dominant feature  at 1.7-2.1\,\AA.
The X-ray  emitting plasma in stellar winds is conventionally modeled 
as a thermal collisional plasma. The spectra of the Arches stars can be 
fitted with temperatures above 2\,keV and high
absorption  column densities (see Table~1). However, the flux predicted
by such models at  the iron K-shell line  is a factor of 3 lower than the
observed flux.  The additional contribution can be attributed to  the
fluorescence of nearly  neutral iron. The presence of the same $\lambda
1.7-2.1$\AA\ emission feature in the  spectra of the diffuse emission from 
the Arches supports this suggestion.  However, the  origin of cool 
fluorescenting material should be further explained. An exemplary fit
with  the {\it apec+gaussian} model to the spectrum of the emission
line star  A2 is shown in Fig.\,4. 

The Arches cluster contains hundreds of massive OB stars (F02), each
of  them expected to be an  X-ray source. Yet only a handful of point
sources is detected within the cluster  radius.   The scarcity of
stellar  X-ray sources in the Arches is naturally explained by the high
interstellar  absorption in the Galactic center. At the column density
of  $\sim\,10^{23}$~cm$^{-2}$ the photoelectric absorption  attenuates
photons up to energies of 2\,keV. Only sources that are sufficiently 
luminous in the  hard X-ray band  (2.5 -- 10 keV) can be observed.

X-ray temperatures of several keV are expected and have been observed
in wide  binary systems (e.g. WR140), where an {\em adiabatic}
colliding wind shock is  expected to occur.  Antokhin et al.\,(2004) 
calculated X-ray spectra of {\it radiative} colliding winds shocks.
They conclude   that the presence of an ``iron bump'' at 6.7\,keV can
also be expected in  relatively close  binary systems (orbital period up 
to the order of 10 days). 

However, by far not all binary stars show hard X-rays (harder than 2\,keV). 
Many X-ray luminous spectroscopic binaries have X-ray temperatures not higher 
than the canonical 0.6\,keV  predicted for single O stars from hydrodynamic
models (Feldmeier et al.\,1997). An interesting example of such binary is 
HD\,150136,  one of the most X-ray luminous O stars (Skinner et al.\,2005). 
The emission from this binary star is dominated by plasma with temperature 
$\approx\,0.3$\,keV, with a small contribution of $\approx\,1$\,keV plasma. 

We employed the model used by Skinner et al.\,(2005) to fit the spectrum of 
HD\,150136, but applied the interstellar absorption and distance appropriate 
for the Arches cluster. The flux of HD\,150136 if it were located in the Arches 
would be at least thousand times smaller, and the system would remain undetected 
by this sensitive Chandra observation. The high interstellar absorption effectively 
selects sources with hard energy distribution. Soft sources, even 
intrinsically luminous, can not be seen in the Galactic center clusters.     

A possible alternative to a colliding wind scenario to explain the hard X-rays 
from the Arches point sources is the generation of X-rays by magnetically confined 
wind flows, as observed in some young stars 
(e.g. $\theta^1$\,Ori\,C, Schulz et al.\,2003). The hardness of the Arches sources 
spectra can be indicative for the presence of stellar or circumstellar 
magnetic fields. A study of the X-ray variability shall shed light on this
proposition. 

\begin{table}
\begin{center}
\begin{tabular}{cccc}    \hline
Object  & Count Rate~[c/s] & $N_{\rm H}$~[cm$^{-2}$]   & $L_{\rm X}$~[erg/s] \\ 
\hline
A1N     &   0.005          & $\approx 7\times 10^{22}$ & $6\times 10^{32}$  \\
A1S     &   0.009          & $\approx 7\times 10^{22}$ & $1\times 10^{33}$  \\
A2      &   0.006          & $\approx 7\times 10^{22}$ & $9\times 10^{32}$  \\
\hline
\end{tabular}
\end{center}
\caption{X-ray emission from point sources in the Arches}
\end{table}
      
The Quintuplet cluster is 1--2 Myr older than the Arches, containing the largest 
conglomerate of WR stars known in the Galaxy (Figer et al.\,1999). It is located 
in a similar environment and at approximately the same distance as the Arches 
(within 50 pc projected distance from Galactic center). Its mass and stellar 
density are comparable with the Arches. Fig.\,2 displays an ACIS-I image of the 
Quintuplet cluster.

There are a few luminous blue variable (LBV) candidates in the Quintuplet. LBVs are 
massive stars in a short phase of large, episodic mass loss. Famous is the Pistol 
star, which is considered to be the most massive star known in the Galaxy. 
To reconcile its prodigious mass estimated as $\sim 150-250 M_\odot$ with the 
recently proposed upper mass cut-off at $150\,M_\odot$ it was suggested that the 
star is a binary. The Pistol star is not detected by the long Chandra exposure, 
implying an upper limit $\log L_{\rm X}/ L_{\rm bol} < -8$. Thus there is no 
support for the binarity assumption from hard X-rays,  which would be an 
indication of colliding stellar winds. 

The scatter of X-ray luminosities among field WR stars is significantly larger 
than among O stars. The X-ray bright WR stars are rare and are binaries 
(sometimes suspected).  Fig.\,2 shows the position of known WR stars in the 
Quintuplet cluster. None of these stars is detected, putting an upper limit on 
their X-ray luminosity at $10^{32}$\,erg/s. The fact that such a young and massive 
star cluster as the Quintuplet is barely visible on the X-ray sky tells that 
the X-ray luminosity of a massive star diminishes significantly when the star 
evolves. 

Among a few weak point sources detected by Chandra, two (FMM\,231 and FMM\,211) 
are members of eponymous quintuplet of cocoon-like objects (Glass et al.\,1999).
The nature of these objects is under debate; their possible identification 
with dusty WC stars has been proposed. However, the X-ray emission from a single 
WC star is highly unusual and is not detected elsewhere. 
  
To conclude: 1) the stellar X-ray sources in the Arches clusters may be attributed 
to colliding wind binaries; 2) the strong emission at 6.4--6.7 keV is observed in 
both stellar and diffuse emission sources in the Arches cluster. The level of this 
emission cannot be reproduced by thermal plasma models. The strong 6.4--6.7 keV 
emission may indicate the presence of cool fluorescenting material in the cluster;
3) the non-detection of X-rays from the Pistol star constrains the hypothesis about
its  binary nature; 4) none of the WR stars known in the Quintuplet cluster is 
detected in X-rays, confirming the general trend of WR stars being X-ray dim.  

%
%
\section*{Acknowledgments}
This research was supported by DFG grant FE 573/3-P.

%
%
 
\beginrefer
\refer  Antokhin I. I, Owocki S. P., Brown J. C., ApJ, 2004, 611, 434

\refer Chevalier R. A. \& Clegg A., 1985, Nat, 317, 44

\refer Feldmeier A., Puls J., Pauldrach A. W. A., 1997, A\&A, 322, 878

\refer Figer D. F., McLean I. S., Morris M., 1999, ApJ, 514, 202 (FMM99)

\refer Figer D. F. et al., 2002, ApJ, 581, 258 (F02) 

\refer Glass I. S. et. al., 1999, MNRAS 304, L10

\refer Law C. \& Yusef-Zadeh F., 2004, ApJ, 611, 858

\refer Oskinova L. M., 2005, MNRAS, 361, 679 

\refer Schild H. et al., 2004, A\&A, 422, 177

\refer Schulz N. S., et al., 2003, ApJ, 595, 365
  
\refer Skinner S. L., et al., 2005, MNRAS, 361,191
  
\endrefer          
\end{document}